\documentclass[twocolumn,showpacs,preprintnumbers,amsmath,amssymb]{revtex4}

\usepackage{graphicx}

\begin{document}

\title{Upper critical field study in the organic superconductor $\beta''$-(ET)$_{2}$SF$_{5}$CH$_{2}$CF$_{2}$SO$_{3}$ : Possibility of Fulde-Ferrell-Larkin-Ovchinnikov state}

\author{K. Cho}
\author{B. E. Smith}
\author{W. A. Coniglio}
\author{L. E. Winter}
\author{C. C. Agosta}
\affiliation{Department of Physics, Clark University, Worcester, Massachusetts 01610}

\author{J. A. Schlueter}
\affiliation{Chemistry and Materials Science Division, Argonne National Laboratory, Argonne, Illinois 60439}

\date{\today}

\begin{abstract}
We report upper critical field measurements in the metal-free-all-organic superconductor $\beta''$-(ET)$_{2}$SF$_{5}$CH$_{2}$CF$_{2}$SO$_{3}$ obtained from measuring the in-plane penetration depth using the tunnel diode oscillator technique. For magnetic field applied parallel to the conducting planes the low temperature upper critical fields are found to exceed the Pauli limiting field calculated by using a semi-empirical method. Furthermore, we found a signature that could be the phase transition between the superconducting vortex state (SCVS) and the Fulde-Ferrell-Larkin-Ovchinnikov (FFLO) state in the form of a kink just below the upper critical field and only at temperatures below 1.23\,K. 
\end{abstract}

\maketitle

In this paper we report the upper critical field $H_{\rm c2}$ versus temperature phase diagram of $\beta''$-(ET)$_{2}$SF$_{5}$CH$_{2}$CF$_{2}$SO$_{3}$ over temperatures ranging from 0.45\,K to 4.0\,K and find an enhancement of $H_{\rm c2}$ over the Pauli limiting field $H_{\rm p}$ at temperatures below 1.23\,K. An additional lower field phase transition which appears as a kink just below $H_{\rm c2}$ and only at temperatures below 1.23\,K has been observed. We suggest that this lower field phase transition could be a signature of the SCVS-to-FFLO phase transition.

Since the first discovery of organic superconductors in 1980~\cite{Jerome}, their unique features, such as low dimensionality and cleaness due to self-organizing, have drawn a lot of interest~\cite{Singleton2}. Due to these features, several theoretical studies have suggested organic superconductors as possible candidates for the FFLO state~\cite{Bulaevskii, Shimahara2, Casalbuoni}.

The FFLO state is an additional superconducting state which stablizes above $H_{\rm p}$ when the superconducting order parameter lowers its free energy by becoming spatially inhomogeneous in a high magnetic field. It was predicted by Fulde and Ferrell~\cite{Fulde} and independently by Larkin and Ovchinnikov~\cite{Larkin} in 1964. Subsequent theoretical studies have shown that heavy fermion and quasi-two-dimensional organic superconductors are possible candidates for the FFLO state~\cite{Bulaevskii, Shimahara2, Casalbuoni}. In the past decade several heavy fermion materials, such as CeRu$_{2}$~\cite{Huxley} and UPd$_{2}$Al$_{3}$~\cite{Gloos}, were suggested to have a FFLO state, but these observations were later found to be due to flux motion~\cite{Tenya} or be inconsistent with theoretical models~\cite{Norman}. A more unambiguous observation of the FFLO state was reported in CeCoIn$_{5}$ by measuring the specific heat and magnetization~\cite{Radovan} and confirmed by subsequent specific heat~\cite{Bianchi} and penetration depth measurements~\cite{Martin}. In organic superconductors there have been several reports of the possible existence of the FFLO state~\cite{Singleton, Tanatar, Uji, Lortz}.

In the metal-free-all-organic superconductor $\beta''$-(ET)$_{2}$SF$_{5}$CH$_{2}$CF$_{2}$SO$_{3}$ the enhancement of $H_{\rm c2}$ above the value of $H_{\rm p}$ has been observed at low temperatures~\cite{Zuo}. This observation raised the possibility of the existence of the FFLO state. The material also satisfies two important conditions necessary to become a candidate for the FFLO state. One is that the material be in the Pauli limit, which means that the orbital limiting field $H_{\rm orb}$ is much greater than $H_{\rm p}$. $H_{\rm orb}$ is the magnetic field above which the Cooper pairs are destroyed when the kinetic energy due to the formation of vortices exceeds the condensation energy of the Cooper pairs. The Pauli limiting field $H_{\rm p}$ is the magnetic field above which the Cooper pairs are destroyed when the Zeeman splitting energy exceeds the condensation energy of the Cooper pairs. These two limiting fields can simultaneously contribute to destroy the Cooper pairs, but under certain circumstances, one mechanism dominates the other. The Pauli limiting condition can dominate when a magnetic field is applied parallel to the conducting planes in a highly anisotropic superconductor~\cite{Shimahara2, Mansky, Burkhardt}. Whether a superconductor is orbitally limited or Pauli limited can be estimated by calculating the ratio of two limiting fields, known as the Maki parameter $\alpha = \sqrt{2} (H_{\rm orb} / H_{\rm p})$~ \cite{Maki2}. For $\beta''$-(ET)$_{2}$SF$_{5}$CH$_{2}$CF$_{2}$SO$_{3}$ with the magnetic field aligned in conducting planes, $H_{\rm orb}$ is calculated to be 39\,T~\cite{Muller} following the Werthamer-Helfand-Hohenberg formula $H_{\rm orb}(T=0) = 0.7T_{c} |d H_{\rm c2}/dT|_{\rm T_{\rm c}}$~\cite{Werthamer}, and $H_{\rm p}$ is 9.9\,T by using a semi-empirical method~\cite{Braunen} with a fit to specific heat data~\cite{Wanka}. From these values $\alpha$ is found to be 3.94 ($ >$1.8), which means that this material is in the Pauli limit~\cite{Gruenberg}. The other condition is that the material be in the clean limit, which is important because the FFLO state is unstable in the presence of normal impurities. According to our upper critical field and Shubnikov-de Haas measurements~\cite{Kyuil}, the coherence length $\xi_{||}$ and mean free path $l_{||}$ for $\beta''$-(ET)$_{2}$SF$_{5}$CH$_{2}$CF$_{2}$SO$_{3}$ are 158\,\AA\,  and 2080\,\AA, respectively, showing that the material is in the clean limit ($l_{||}$/$\xi_{||}$ $>1$).

\begin{figure}
\includegraphics{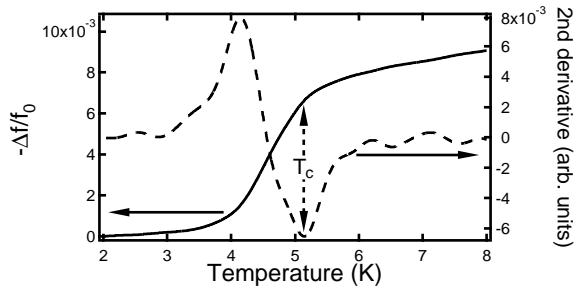}
\caption{\label{fig:fig1_Tc} (Sample A) Solid line (left axis) is the temperature dependence of normalized frequency change (-$\Delta f / f_0$). The dotted line (right axis) is the second derivative of the left axis with respect to temperature. The minimum of the second derivative is defined as $T_{\rm c}$ = 5.15\,K in this case.}
\end{figure}

Single crystals of $\beta''$-(ET)$_{2}$SF$_{5}$CH$_{2}$CF$_{2}$SO$_{3}$ were synthesized by the electrocrystallization technique~\cite{Geiser}. The dimensions of the two samples used in this experiment were $735 \times 490 \times 245\,\mu$m$^{3}$ (sample A) and $813 \times 762 \times 432\,\mu$m$^{3}$ (sample B). We measured the in-plane radio frequency (rf) penetration depth with 35\,MHz (sample A) and 63\,MHz (sample B) tunnel diode oscillators (TDO)~\cite{Coffey} by putting the conducting planes of the sample perpendicular to the axis of the TDO inductor coil. To orient the conducting planes of the sample accurately with respect to the applied magnetic field, a one axis rotating probe with a 0.23$^{\circ}$ angular resolution was used. The temperature was swept from 0.45\,K to 8.0\,K. The applied field was created with a 43\,T pulsed magnet with a 25\,ms rise-time. For each temperature several pulsed shots at slightly different angle increments were used to find $H_{\rm c2}$.  To reduce the induced voltage during a pulsed field shot, a balanced counter-wound TDO coil was used. The background data without a sample was measured at the identical conditions and subtracted.

Measurement of the rf penetration depth with a TDO is an accurate probe to detect $H_{\rm c2}$. During a magnetic field pulse, a change of the in-plane rf penetration depth produces a small enough change in the TDO frequency that there is a linear relation between the changes of frequency ($\Delta f$) and in-plane rf penetration depth ($\Delta \lambda$) as $\Delta \lambda = - \Gamma \Delta f$, where $\Gamma$ is a geometrical factor calculated from the sample and the coil. The phase transition between superconducting and normal states can be accurately observed given the high signal to noise ratio during the transition. For our purpose of locating $H_{\rm c2}$, we will directly use the frequency change instead of converting it to penetration depth. 

\begin{figure}
\includegraphics[scale=0.75]{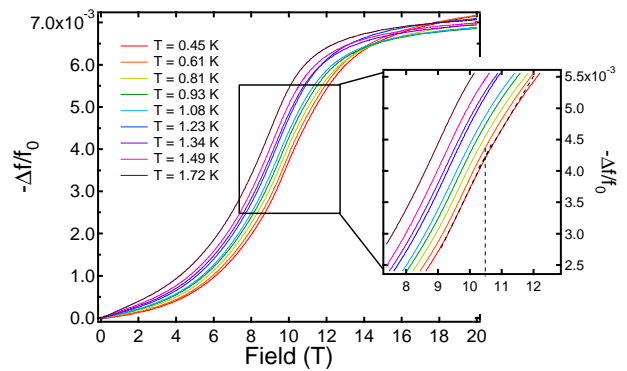}
\caption{\label{fig:fig2_ParallShots} (Color online) (Sample A) Normalized frequency changes (-$\Delta f/f_0$) during pulsed field shots at low temperatures when applied fields are parallel to the conducting planes of the sample. $f_{0}$ is an initial TDO frequency at zero field. Background subtracted. Note that an additional transition is observed below 1.23\,K as a kink. The inset is a blow-up of the area of the kink. The dotted line in the inset shows a well-defined kink at 0.45\,K.}
\end{figure}

Figure~\ref{fig:fig1_Tc} shows the temperature dependent frequency change at zero applied field for sample A. From this plot we find the critical temperature $T_{\rm c}$ to be 5.15\,K from the second derivative minimum of frequency change (-$ \Delta f/f_0$). The $T_{\rm c}$ measured here is very close to the value determined by penetration depth measurements conducted by Prozorov \textit{et al.}~\cite{Prozorov}. 

\begin{figure}
\includegraphics{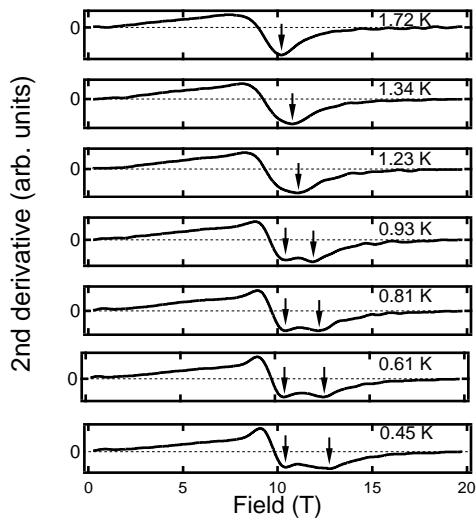}\caption{\label{fig:fig3_2ndDeriv} (Sample A) Second derivatives of Fig.~\ref{fig:fig2_ParallShots}. Only one minimum appears above 1.23\,K but two minima below 1.23\,K. We designate the higher field minimum as $H_{\rm c2}$ and suggest that the lower field minimum is a signature of the SCVS-to-FFLO phase transition.}
\end{figure}

Figure~\ref{fig:fig2_ParallShots} shows the pulsed field shots for sample A at low temperatures. In the plot, the SCVS-to-normal state phase transition is very broad, making it difficult to define $H_{\rm c2}$. Therefore, we used the second derivative method to define $H_{\rm c2}$, as we did to define $T_{\rm c}$ in Fig.~\ref{fig:fig1_Tc}. These upper critical fields will be shown in Fig.~\ref{fig:fig5_Hc2}. In addition and more interestingly, we see in Fig.~\ref{fig:fig2_ParallShots} that weak kinks emerge at around 10.5\,T and only at temperatures below 1.23\,K. The dotted lines in the inset give an example of a well-defined kink at 0.45\,K. These kinks are more obvious in the derivatives that will be shown in Fig.~\ref{fig:fig3_2ndDeriv}. 

\begin{figure}
\includegraphics{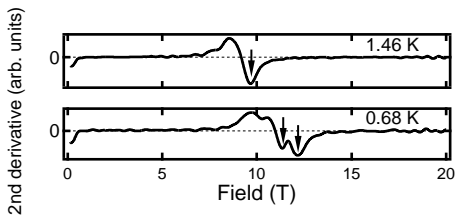}
\caption{\label{fig:fig4_2ndDeriv_sampleB} (Sample B) Second derivatives versus field. Only one minimum appears at 1.46\,K but two minima  at 0.68\,K. The appearance of the lower field additional minimum is consistent with the result for sample A.}
\end{figure}

\begin{figure}
\includegraphics[scale=0.95]{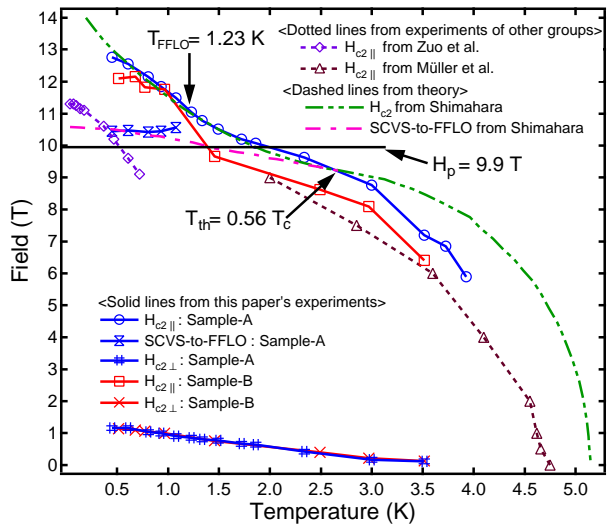}
\caption{\label{fig:fig5_Hc2} (Color online) $H_{\rm c2}$ versus $T$ phase diagram. The circle solid line and bowtie solid line are $H_{\rm c2}$ and lower field phase transition for sample A measured for magnetic field parallel to the conducting planes, respectively. The square solid line is $H_{\rm c2}$ for sample B. The diamond dotted line and triangle dotted line are $H_{\rm c2}$ from the resistivity measurement in Ref.~\cite{Zuo} and the thermal coefficient measurement in Ref.~\cite{Muller}. The pound solid line and cross solid line are $H_{\rm c2}$ of samples A and B for magnetic field perpendicular to the conducting planes. Dashed lines are the theoretical calculation from Shimahara~\cite{Shimahara1994}. The double dot dashesd line is for $H_{\rm c2}$ and the single dot dashed line for the SCVS-to-FFLO phase transition. The theoretical curves are scaled by using $H_{\rm c2}(T=0)$ = 15\,T and $T_{c}$ = 5.15\,K.}
\end{figure}

Figure~\ref{fig:fig3_2ndDeriv} plots the low temperature second derivatives corresponding to the pulsed field shots in Fig.~\ref{fig:fig2_ParallShots}. For each temperature above 1.23\,K only one minimum is clearly observed, which can be easily assigned as $H_{\rm c2}$. For each temperature below 1.23\,K, however, two minima are observed. The higher field minimum corresponds to $H_{\rm c2}$ at a given temperature because it is consistent with the top inflection area in Fig.~\ref{fig:fig2_ParallShots} and continuous from higher temperatures above 1.23\,K. The lower field minimum observed below 1.23\,K corresponds to the kink at a given temperature in Fig.~\ref{fig:fig2_ParallShots}. This lower field minimum will be considered as a signature of the SCVS-to-FFLO phase transition in the following. We obtained a similar result with sample B, although the data was noisier. In Fig.~\ref{fig:fig4_2ndDeriv_sampleB} two minima are clearly observed at 0.68\,K, and only one minimum is observed at 1.46\,K.

All the minima of samples A and B that are obtained from the second derivative method are plotted in Fig.~\ref{fig:fig5_Hc2} and compared with the results by M\"uller \textit{et al.}~\cite{Muller} and Zuo \textit{et al.}~\cite{Zuo}. The upper critical fields for samples A (circle solid line) and B (square solid line) are consistent with each other over the whole temperature range. The high temperature regions above 1.4\,K for samples A and B have a similar curvature and are consistent with M\"uller \textit{et al.}'s thermal coefficient measurement (triangle dotted line)~\cite{Muller}. The low temperature regions below 1.23\,K for samples A and B both show that $H_{\rm c2}$ increases above $H_{\rm p}$ with decreasing temperature. This increasing behavior of $H_{\rm c2}$ is qualitatively consistent with Zuo \textit{et al.}'s resistivity measurement (diamond dotted line)~\cite{Zuo}, even though there is a quantitative difference by almost two tesla partially due to the difference of the measurement methods. We found that both experiments on samples A and B show a clear slope change at around 1.23\,K. 

The additional lower field phase transition of sample A which appears below 1.23\,K in Fig.~\ref{fig:fig3_2ndDeriv} is plotted as a bowtie solid line in Fig.~\ref{fig:fig5_Hc2}. This lower field transition line is located between $H_{\rm c2}$ and $H_{\rm p}$. More interestingly, the lower field transition starts almost at the same temperature as the strong slope change of $H_{\rm c2}$ (circle solid line) occurs, which is also 1.23\,K. The coincidence between the slope change of $H_{\rm c2}$ at 1.23\,K and the onset of the lower field transition at 1.23\,K strongly suggests that the enhancement of $H_{\rm c2}$ at low temperatures is related to the additional lower field transition, that is, the possible existence of the FFLO state. Here, we define $T_{\rm FFLO}$ = 1.23\,K (= 0.24\,$T_{\rm c}$).

The phase diagram of the SCVS-to-FFLO state (bowtie solid line) and the FFLO-to-normal state(circle solid line) phase transitions of sample A in Fig.~\ref{fig:fig5_Hc2} is very consistent with the theoretical expectation (single-dot and double-dot dashed lines) calculated by Shimahara~\cite{Shimahara1994}. $T_{\rm FFLO}$ of $1.23\,K$ is 0.24\,$T_c$ smaller than the theoretical upper limit of $T_{th} = 0.56\,T_{c}$. In comparison with the previous results in organic and heavy fermion superconductors~\cite{Radovan, Bianchi, Martin, Singleton, Tanatar}, the location of the SCVS-to-FFLO phase transition (bowtie solid line) in Fig.~\ref{fig:fig5_Hc2} is topologically identical to them. For the organic superconductors, a vortex stiffness measurement in $\kappa$-(ET)$_{2}$Cu(NCS)$_{2}$~\cite{Singleton} shows that $T_{\rm FFLO}=0.56\,T_{\rm c}$, and a thermal conductivity measurement in $\lambda$-(BETS)$_{2}$GaCl$_{4}$~\cite{Tanatar} gives $T_{\rm FFLO}=0.35\,T_{\rm c}$. Both results are greater than our result where $T_{\rm FFLO}= 0.24\,T_{\rm c}$, but are smaller than the theoretical upper bound of $T_{\rm th} = 0.56\,T_{\rm c}$~\cite{Gruenberg, Agterberg, Shimahara1994}. Also, the SCVS-to-FFLO phase transition (bowtie solid line) is almost constant below $T_{\rm FFLO}$, which is similar to the result of thermal conductivity measurement in $\lambda$-(BETS)$_{2}$GaCl$_{4}$~\cite{Tanatar}. However, for the vortex stiffness measurement in $\kappa$-(ET)$_{2}$Cu(NCS)$_{2}$, the SCVS-to-FFLO phase transition increases as the temperature decreases from $T_{\rm FFLO}$ \cite{Singleton}. For CeCoIn$_{5}$ the values of $T_{\rm FFLO}$ from specific heat~\cite{Radovan, Bianchi} and penetration depth measurements~\cite{Martin} are consistent with each other and give $0.14\,T_{c}$, which is much smaller than our result. Unlike the case of organic superconductors, the SCVS-to-FFLO phase transition in CeCoIn$_{5}$ decreases in field as the temperature decreases. 

The TDO technique measures the in-plane penetration depth ($\lambda$) which is sensitive to the density of superconducting electrons ($n_s$)~\cite{Tinkham}. If the superconducting phase changes from the SCVS to the FFLO state, the corresponding change in $n_s$ induces the variation in $\lambda$. As seen in Fig.~\ref{fig:fig2_ParallShots}, the broad first-order phase transition ($H_{\rm c2}$) seems to include the kink-like second-order SCVS-to-FFLO phase transition. Hence, it is hard to define the order of the phase transitions. Theoretically, the phase trasition is predicted to be a second-order for the SCVS-to-FFLO and first-order for the FFLO-to-normal~\cite{Houzet, Ikeda}. These predictions were confirmed by specific heat measurements of CeCoIn$_{5}$~\cite{Radovan}. In Fig.~\ref{fig:fig2_ParallShots}, we find qualitative consistency in the shape and order of the phase transitions with the theoretical expectations~\cite{Shimahara1994}. To determine the nature of the phase transitions in detail requires more direct thermodynamic measurements.

\begin{figure}
\includegraphics{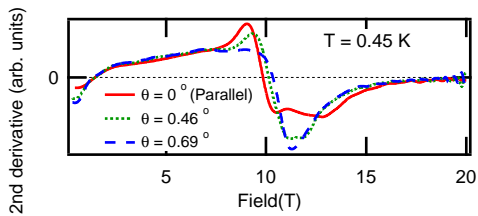}
\caption{\label{fig:fig6_Angle_FFLO} (Color online) (Sample A) Second derivatives versus field at different angles. Two minima shown at the parallel angle become one as the angle gets farther off the parallel.}
\end{figure}

In Fig.~\ref{fig:fig6_Angle_FFLO}, the angular dependence of two minima at 0.45\,K is displayed. At the parallel angle ($\theta$ = 0$^{\circ}$), the two minima are clearly observed. As the angle gets farther from parallel, the two minima become one at about 0.7$^{\circ}$. This strongly suggests that the FFLO state in this material exists only when the magnetic field is very closely aligned with the conducting planes~\cite{Correa}. 

One could argue that the SCVS-to-FFLO phase transition we found could be related to vortex formation. However, as we mentioned, the applied magnetic fields were parallel to the conducting planes of the samples, so the formation and effects of vortices are mostly suppressed~\cite{Martin2}. Moreover, the SCVS-to-FFLO phase transitions are observed in both up and down sweeps of our pulsed field shots at the same field. Thermal cycling during the experiments didn't affect the results. Thus, we saw no hysteretic effects. These facts suggest that our new phase transition is not due to vortices.

In conclusion, we have reported in-plane rf penetration depth measurements in the metal-free-all-organic superconductor $\beta''$-(ET)$_{2}$SF$_{5}$CH$_{2}$CF$_{2}$SO$_{3}$. For magnetic field applied parallel to the conducting planes, we confirmed that $H_{\rm c2}$ for temperatures below 1.23\,K clearly exceeds $H_{\rm p}$ and increases with decreasing temperature. In particular, a slope change of $H_{\rm c2}$ at around 1.23\,K was found and coincided with the onset of a new lower field phase transition. This observation could be a signature of the SCVS-to-FFLO phase transition. To verify its origin, more experiments are required.

\begin{acknowledgments}
The work is supported by the U.\ S.\ Department of Energy, BES, under Grant No.\ DE-FG02-05ER46214.
\end{acknowledgments}

\end{document}